\documentclass[useAMS,usenatbib]{mn2e}

\usepackage{graphicx}

\title[CHIMERA: Caltech HIgh-speed Multi-color camERA]{CHIMERA: a wide-field, multi-color, high-speed photometer at the prime focus of the Hale telescope}
\author[L. K. Harding et al.]{L. K. Harding,$^{1,2}$\thanks{Instrument Scientist; E-mail: leon.k.harding@jpl.nasa.gov} G. Hallinan,$^{1}$\thanks{Principal Investigator; E-mail: gh@astro.caltech.edu} J. Milburn,$^{1}$ P. Gardner,$^{1}$ N. Konidaris,$^{1}$ \newauthor N. Singh,$^{1,2}$ M. Shao,$^{2}$ J. Sandhu,$^{2}$ G. Kyne,$^{1}$ and H. E. Schlichting$^{3}$
 \\
$^{1}$Cahill Center for Astronomy \& Astrophysics, California Institute of Technology, Pasadena CA 91125, USA\\
$^{2}$Jet Propulsion Laboratory, California Institute of Technology, Pasadena CA 91109, USA\\
$^{3}$Massachusetts Institute of Technology, 77 Massachusetts Avenue, Cambridge, MA 02139-4307, USA}
\begin{document}

\date{Accepted 2016 Jan 11. Received 2016 Jan 06; in original form 2015 Jun 09}

\pagerange{\pageref{firstpage}--\pageref{lastpage}} \pubyear{0000}

\maketitle

\label{firstpage}

\begin{abstract}
The Caltech HIgh-speed Multi-color camERA (CHIMERA) is a new instrument that has been developed for use at the prime focus of the Hale 200-inch telescope.  Simultaneous optical imaging in two bands is enabled by a dichroic beam splitter centered at 567 nm, with Sloan u$^{\prime}$ and g$^{\prime}$ bands available on the blue arm and Sloan r$^{\prime}$, i$^{\prime}$ and z\_s$^{\prime}$ bands available on the red arm.  Additional narrow-band filters will also become available as required.  An Electron Multiplying CCD (EMCCD) detector is employed for both optical channels, each capable of simultaneously delivering sub-electron effective read noise under multiplication gain and frame rates of up to 26 fps full frame (several 1000 fps windowed), over a fully corrected 5 $\times$ 5 arcmin field of view.  CHIMERA was primarily developed to enable the characterization of the size distribution of sub-km Kuiper Belt Objects via stellar occultation, a science case that motivates the frame-rate, the simultaneous multi-color imaging and the wide field of view of the instrument.  In addition, it also has unique capability in the detection of faint near-Earth asteroids and will be used for the monitoring of short duration transient and periodic sources, particularly those discovered by the intermediate Palomar Transient Factory (iPTF), and the upcoming Zwicky Transient Facility (ZTF).

\end{abstract}


\section{Introduction}

The Kuiper Belt consists of a disk of icy bodies located beyond the orbit of Neptune.  Determining the abundance, material properties, and collisional processes of sub-km-sized Kuiper Belt Objects (KBOs) is important, since these bodies provide the link between the largest KBOs and the dust-producing debris disks observed around other stars.  Observations and theory suggest the existence of a break in the power-law size distribution at smaller KBO radii (\citet{schlichting12}, and references therein).  The break in the size distribution is generally attributed to collisions that break-up small KBOs (radius $r \leq$ 10 km) and that therefore modifies their size distribution.  Importantly, this interpretation has not been tested observationally, as KBOs $<$10 km in radius are too faint to be in detected in reflected light. 

However, such objects can be detected indirectly by stellar occultations.  A small KBO crossing the line of sight of a star will partially obscure the star. For sub-km-sized KBOs, diffraction effects become important and the duration of the occultation is approximately given by the ratio of the Fresnel scale to the relative velocity perpendicular to the line of sight between the observer and the KBO. The Fresnel scale is given by $\sqrt{\lambda / 2 \cdot a} \sim 1.3$ km, where $a\sim 40$\,AU is the distance to the Kuiper Belt, and $\lambda \sim 600$ nm is the wavelength of the observation.  Since the relative velocity is usually dominated by the Earth's velocity around the Sun, which is \mbox{30 km sec$^{-1}$}, typical occultations only last of order of a tenth of a second.

To date, only a very small number of high significance detections of KBO occultations have been reported.  Most notably, 31,500 star hours of archival data from the Hubble Space Telescope (HST) Fine Guidance Sensors (FGS)  recorded with 40 Hz sampling frequency, was used to search for stellar occultations by small KBOs \citep{schlichting09, schlichting12}.  In order to increase the number of detections significantly, ground-based surveys for similar occultation events are required. However, such a program presents unique observational challenges, specifically: 

\begin{enumerate}

\item Statistics from HST data suggest that a large number of stars need to be monitored simultaneously in order to achieve a sufficient number of star hours to significantly improve on the existing sample.
\item Unlike the KBO occultations found in HST FGS data, robust means to rule out false positives caused by atmospheric scintillation noise is essential for ground-based searching for KBO occultations.
\item A high sampling frequency of 40 Hz is required such that one can adequately resolve the diffraction pattern of the occultation event.

\end{enumerate}

CHIMERA was conceived, designed and built to search for sub-km sized KBOs in the outer solar system, by detection of stellar occultations over a wide field of view (FOV).  Our design was optimized to address the specific challenges outlined above, in that: 

\begin{enumerate}

\item The number of stars that can be simultaneously monitored for KBO occultations scales with the areal FOV. We use the largest format Electron Multiplying CCDs (\mbox{EMCCDs}) currently available (1024 $\times$ 1024 pix; see Section~\ref{EMCCDs}), which, together with the additional requirement of adequate sampling of the seeing-limited point spread function (PSF) for the best seeing observed at Palomar (\mbox{$\sim$0.7 arcsec}), limits our FOV to 5 $\times$ 5 arcmin. A future generation of CHIMERA is planned to be developed in parallel with larger format ($>$4K $\times$ 4K) EMCCD and sCMOS sensors and will thus increase the areal FOV substantially. Nonetheless, even with the existing FOV, CHIMERA can monitor thousands of stars simultaneously with sufficient signal to noise in a 25 ms integration to search for the signature of a KBO occultation (see first light image, Figure~\ref{fig9}, \textit{right}). This is achieved by targeting dense star fields in regions of the sky where the ecliptic and galactic planes overlap.

 \item Accounting for false positives due to extreme scintillation is perhaps the most difficult challenge facing ground-based KBO occultation detection efforts. The Transneptunian Automated Occultation Survey (TAOS II) will make use of three 1.3\,m telescopes, requiring simultaneous detection in each telescope to rule out scintillation \citep{lehner14}. By availing of the 5.1\,m aperture of the Hale telescope, CHIMERA will exhibit lower scintillation noise (function of D$^{-2/3}$) than searches on small telescopes. Furthermore, the dense fields chosen for KBO occultation searches will help us further in this regard, in that a large reduction in the scintillation noise power can be achieved through differential photometry using groups of nearby stars \citep{kornilov12}. More important, however, is CHIMERA's ability to conduct simultaneous imaging in two photometric bands. The diffraction pattern signature caused by the occultation of a star by a KBO has a specific wavelength dependence that will allow us to differentiate between true occultation events and false positives due to extreme scintillation events, the latter of which will not display the same color-dependent effect \citep{kornilov11}. 

\item The readout speed and noise characteristics of a detector capable of delivering full-frame imaging at 40\,Hz plays a key role in sensitivity to KBO occultation events.  Traditionally, photometers and other similar instruments have used CCD\footnotemark[1] or SCD\footnotemark[1] sensors, or in some cases CMOS\footnotemark[1] or MCP\footnotemark[1] architectures (e.g. \citet{dhillon07,law06,matt10,odonoghue95,stover87,wilson03}, and references therein). \footnotetext[1]{Charge-Coupled Devices (CCD); Segmented Charge-Coupled Devices (SCD); Complementary metaloxidesemiconductor (CMOS); Microchannel plate (MCP)}CCDs have been the preferred detector for the focal planes of telescopes or indeed for photometric instruments; however, they can suffer from large amounts of output amplifier noise, especially at higher read out rates.  CMOS noise can be more difficult to characterize than CCD noise because of additional pixel and column amplifier noise, as well as non-linearities in the charge-to-voltage conversions \citep{holst11}.  With the advent of the EMCCD at the beginning of the last decade \citep{jerram01}, extremely low noise ($<$1 e$^{-}$ rms), high-speed clocking, has become possible.  The architecture of an EMCCD is very similar to that of a CCD.  The difference lies in the EMCCD's so-called high gain, or electron multiplication (EM) register, which is an additional stage containing a large responsivity output.  In this region, electrons can be amplified by a process known as avalanche multiplication.  The result is a much higher signal to noise ratio (S/N), albeit at the proportional reduction in pixel charge capacity. We have taken advantage of the progress in EMCCD detector development to allow us to deliver 40\,Hz imaging across our full field with an effective read noise of $<$1 e$^{-}$ rms under EM gain. 

\end{enumerate} 

A detailed discussion of the optical and mechanical design and construction, as well as the detectors and software development, and instrument performance, involved in delivering the CHIMERA instrument to the specifications described above are presented in Sections 2 -- 6.  Examples of first light data and a brief description of CHIMERA's future plans are described in Section~\ref{section_future}.

\subsection{Additional Key Science}\label{NEA}


\begin{figure*}
   \centering
   \includegraphics[width=15.5cm]{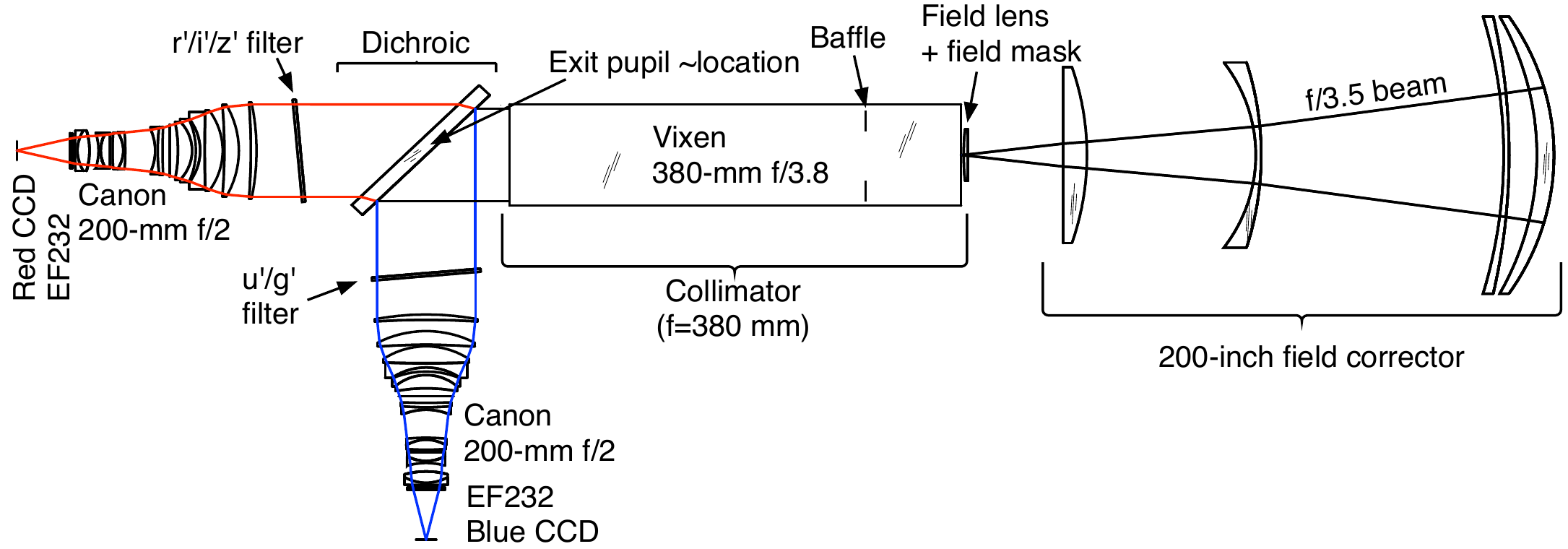}
\caption{Section view and ray trace of the CHIMERA optical layout.  The beam enters from the Hale telescope (right), passes through the 200-inch field corrector, the field lens, field baffle (and stop), entrance baffle, Vixen collimator (mounted backwards), and 45$^{\circ}$ dichroic beam splitter.  At the dichroic, the beam is sent to either the blue or red side.  These paths are both serviced by a Canon 200-mm \textit{f/2.0} lens and EMCCD.  Custom filter exchangers are placed in the collimated beam before the Canon lens, where the blue side currently houses Sloan u$^{\prime}$ or g$^{\prime}$ bands, and the red side either of the Sloan r$^{\prime}$, i$^{\prime}$ or z\_s$^{\prime}$ bands.}\label{fig1}
\end{figure*}

The unique capabilities of CHIMERA, as designed to target KBO occultations, are also well suited for a wide swath of additional high-time resolution astrophysics.  Indeed, the utility of multi-color, high-speed photometers has been amply demonstrated by the ULTRACAM instrument \citep{dhillon07}. CHIMERA can be used to monitor short duration transient and periodic sources, such as aurorae on brown dwarfs, eclipsing binaries, flaring stars, pulsing white dwarfs, and transiting planets, and will be well placed for follow-up observations of short-duration transients discovered by the intermediate Palomar Transient Factory (iPTF), and the upcoming Zwicky Transient Facility (ZTF).  

Additional key science cases will also take advantage of CHIMERA's wide field, particularly searches for near Earth asteroids (NEAs). To date, over 10,000 NEAs have been discovered where 1,000 were $>$1 km in size (see \citet{kaiser02,larson06}). Much like the detection of small mass KBOs, the detection rates for small NEAs (\mbox{$<$50 m}) have been largely unsuccessful, with as much as 98\% of the estimated half a million 50 m-class NEAs not yet discovered \citep{shao14}.  Finding and tracking \mbox{$5 - 10$ m} NEAs presents a major challenge due to how dim they are (H = $28 - 30$ mag), and also because they also exhibit high apparent proper motion (0.14 arcsec sec$^{-1}$ at 0.1 AU).  Traditional surveys operating at $>$30 sec were much less sensitive to fast moving NEAs.  The smallest NEAs discovered have typically been lost thereafter, and if sufficiently accurate astrometric data are not collected, future positions ($4 - 6$ yr) when passing close to Earth cannot be calculated.

A new technique has been recently developed at the Jet Propulsion Laboratory (JPL), \textit{synthetic tracking}, which was designed to find very small NEAs (H=30 mag) -- see \citet{shao14} for full details.  Although the S/N of a single short exposure is insufficient to detect these objects in one frame, by shifting successive frames with respect to each other and subsequently coadding in post-processing, a long-exposure image is synthetically created.  The resulting frame appears as if the telescope were tracking the NEA with $>>$S/N.

CHIMERA's performance, in conjunction with the effectiveness of the \textit{synthetic tracking} technique, is an efficient NEA detection machine.  Automated dithering routines, discussed later in Section~\ref{software}, have been specially developed to conduct blind searches of NEAs over a large fraction of the sky each night.  The limiting magnitude for CHIMERA is \mbox{$\sim$27 mag} (see Section~\ref{performance}), and thus the expected yield for NEA observations with CHIMERA is predicted to be 8 NEAs ($<$10 m, H $ = 28$ mag) and 20 (H$>$25 mag) per night, with sizes down to 7 m, which dwarfs the total global yield of 30 NEAs yr$^{-1}$ for H$>$28 mag.  This prediction has already been demonstrated with the detection of one new low mass NEA with a diameter of 8 m and H $= 29$ mag with an early prototype version of the CHIMERA instrument \citep{shao14, zhai14}.

\section[]{Optical Design}\label{optical_design}

The CHIMERA science objectives imposed strict requirements on the optical design in order to provide: \textit{i)} the capability to image simultaneously in two optical bands, \textit{ii)} a large FOV ($\geq$5 $\times$ 5 arcmin) and \textit{iii)} seeing-limited image quality (typically $\sim$1.2 arcsec median seeing at Palomar, yielding a pixel scale of $\sim$0.3 arcsec pix$^{-1}$).

To meet these requirements, we needed to consider the challenges associated with placing an instrument at the prime focus of the Hale 200-inch.  We sought commercial off-the-shelf (COTS) optics that were well-matched to delivering a wide FOV at the 200-inch prime focus.  Consequently, the decision was made to design a collimator-camera system and place CHIMERA behind the Wynne corrector \citep{wynne67}, which provides a 25 arcmin optically flat field.  This design is illustrated in Figure~\ref{fig1}, and described herein.

Although the collimated beam size of 100 mm was too large for COTS dichroic and filter components, the CHIMERA collimating and re-imaging optics were successfully designed using COTS elements.  The collimator consists of a Vixen 380 mm focal length astrograph telescope from Vixen Optics,\footnotemark[2] \footnotetext[2]{http://www.vixenoptics.com/}which is mounted backwards in the instrument.  Vixen provided an encrypted optical prescription via the Zemax\footnotemark[3] \footnotetext[3]{Zemax is an optical design program; https://www.zemax.com/home}``black box'' model system.  We designed a custom dichroic beam splitter which was procured from Custom Scientific, Inc\footnotemark[4]\footnotetext[4]{www.CustomScientific.com}.  The re-imaging camera was selected to be the Canon 200 mm lens, based on its large \textit{f/2.0} aperture, and was simulated using a prescription in the Japanese patent literature (JP 2011, 253050, A).  The Wynne corrector was modeled from the prescription of \citet{kells1998}.  These models allowed us to simulate, and accurately model, the entire optical system in Zemax and predict optical performance, as well as pupil placement.  We show the theoretical throughput from these models in Table~\ref{table1}, and compare the instrument throughput as measured on-sky, in Section~\ref{performance}.

\subsection{Collimator, field lens \& baffle system}

The Vixen telescope is mounted backwards in order to act as a collimator, and is designed with a stop at the first element, such that the entrance pupil sits on the apex of the objective, and where the exit pupil is located $\sim$360 mm in front of the telescope's focal plane.  Since the 200-inch and the Vixen do not have the same pupil position, we included a 500 mm focal length \textit{f/10} Plano-Convex positive field lens (Edmund Optics \#47-397-INK) to rectify this mismatch.  The field lens places the exit pupil at $\sim$25 mm past the stop of the Vixen collimator.  A second baffle is placed on the optical path 101.6 mm after the field mask, oversized by 15\% to \mbox{66.93 mm} to avoid vignetting.  This 2-element baffle configuration acts to reduce any scattered light from the Wynne corrector, where the reduction from element one was estimated to be 98\%.  Although an ideal field lens would place the exit pupil directly at the stop, the cost/lead time-to-throughput gain was not significant enough to consider a custom design.  The field mask was oversized by 15\% to \mbox{42.75 mm} with respect to CHIMERA's full field (corresponding to a converging beam size of \mbox{37.16 mm}), to avoid vignetting and to minimize alignment error.

Based on the CHIMERA optical requirements, the images delivered by the field corrector, field lens, collimator and telescope, when imaged with an ideal camera are \mbox{0.3 arcsec} in rms diameter over 95\% of the field, averaged over four equal area fields and wavelengths from \mbox{400 -- 900 nm}.  The geometric throughput of the collimator is reduced by the f/\# mismatch between the 200-inch telescope (\textit{f/3.5}) and collimator (\textit{f/3.8}).  The internal fresnel losses were calculated to be $\sim$15\% end-to-end, yielding a total throughput of the collimator to be of order 77\%. 

\begin{figure}
\hspace{-3em}
   \includegraphics[width=10cm]{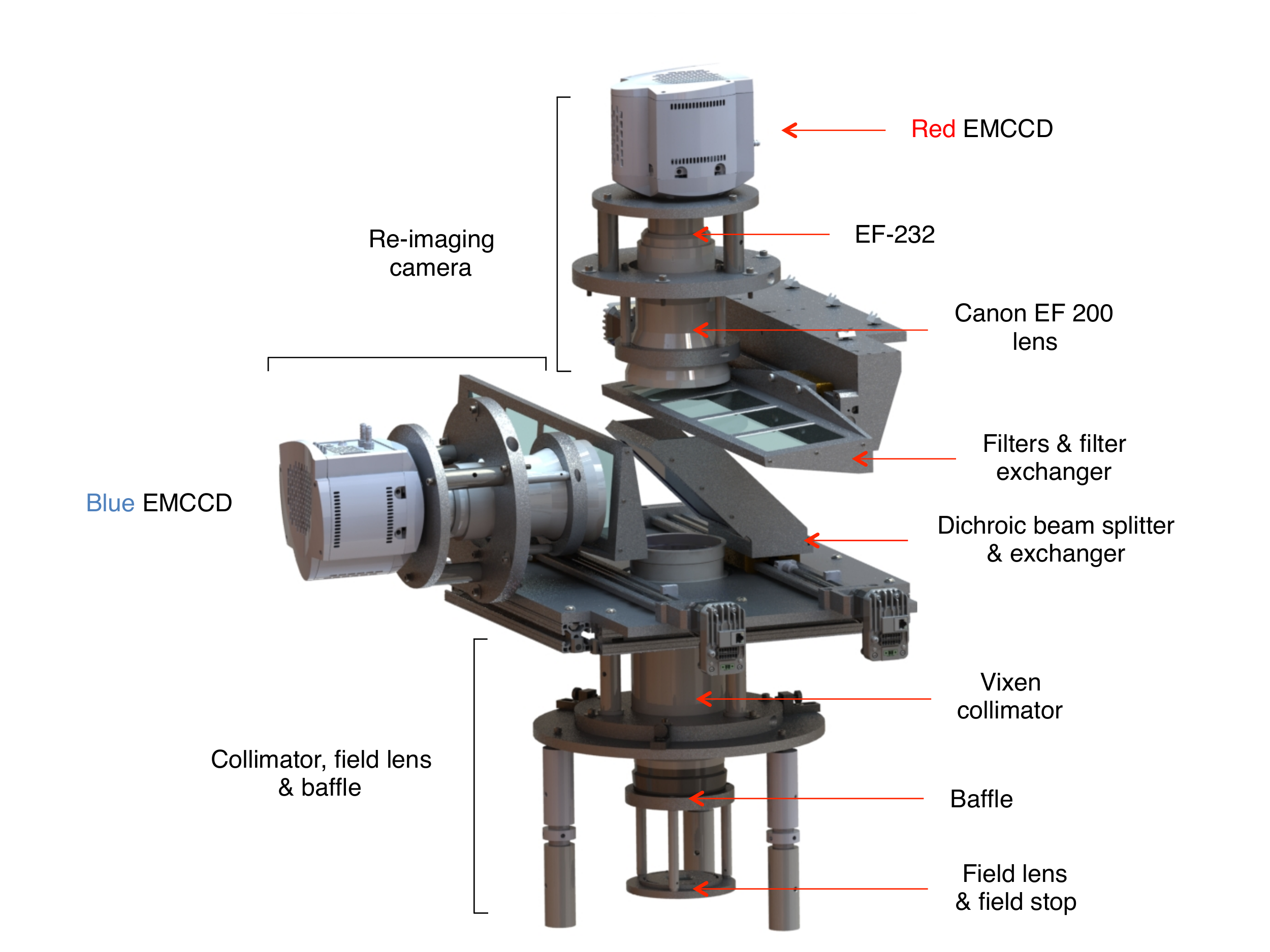}
\caption{3D CAD Solidworks render of the CHIMERA collimator-camera system with its outer structure removed.  This figure illustrates where the optical elements in Figure~\ref{fig1} were integrated in to the instrument's internal space.  The total field of the Vixen collimator is significantly larger than what is used in CHIMERA, therefore the dichroic is held by a black frame covered in felt to prevent scattered light entering the Canon lens.}\label{fig2}
\end{figure}

\subsection{Dichroic beam splitter}

We designed a dichroic beam splitter since an COTS solution was not available that could accommodate the 100 mm unvignetted beam produced by the collimator.  Due to CHIMERA's internal mechanical constraints, the dichroic was positioned 95.25 mm from the Vixen objective, see Figure~\ref{fig2}.  The surface quality and specifications of the dichroic were constrained based on simulations of the telescope, Wynne corrector, collimator and camera system.  This is discussed in more detail in Section 4.

The dichroic element consists of a fused Silica substrate with optical interference coatings on both sides of the glass.  In order to avoid vignetting, we specified the unmounted dimensions to be 111.0 $\times$ 157.0 $\pm$ 0.5 mm, with a minimum clear aperture of 100.0 $\times$ 142.0 $\pm$ 0.5 mm.  The longer dimension was important since the dichroic was designed to accept and split an incoming beam at a 45$^{\circ}$ angle of incidence (AOI), where the central wavelength was at 567 nm.  The transmitted beam has a range of 570 -- 850 nm with a throughput of 90\%, and the reflected beam has a range of 300 -- 540 nm, with an identical throughput to its counterpart.  Careful consideration was given to the dichroic beam splitter's central wavelength and cross-over range of 30 nm, in order to ensure $>$90\% dichroic transmission for each of the adjacent g$^{\prime}$ and r$^{\prime}$ filters.

\begin{figure*}
   \centering
   \includegraphics[width=15cm]{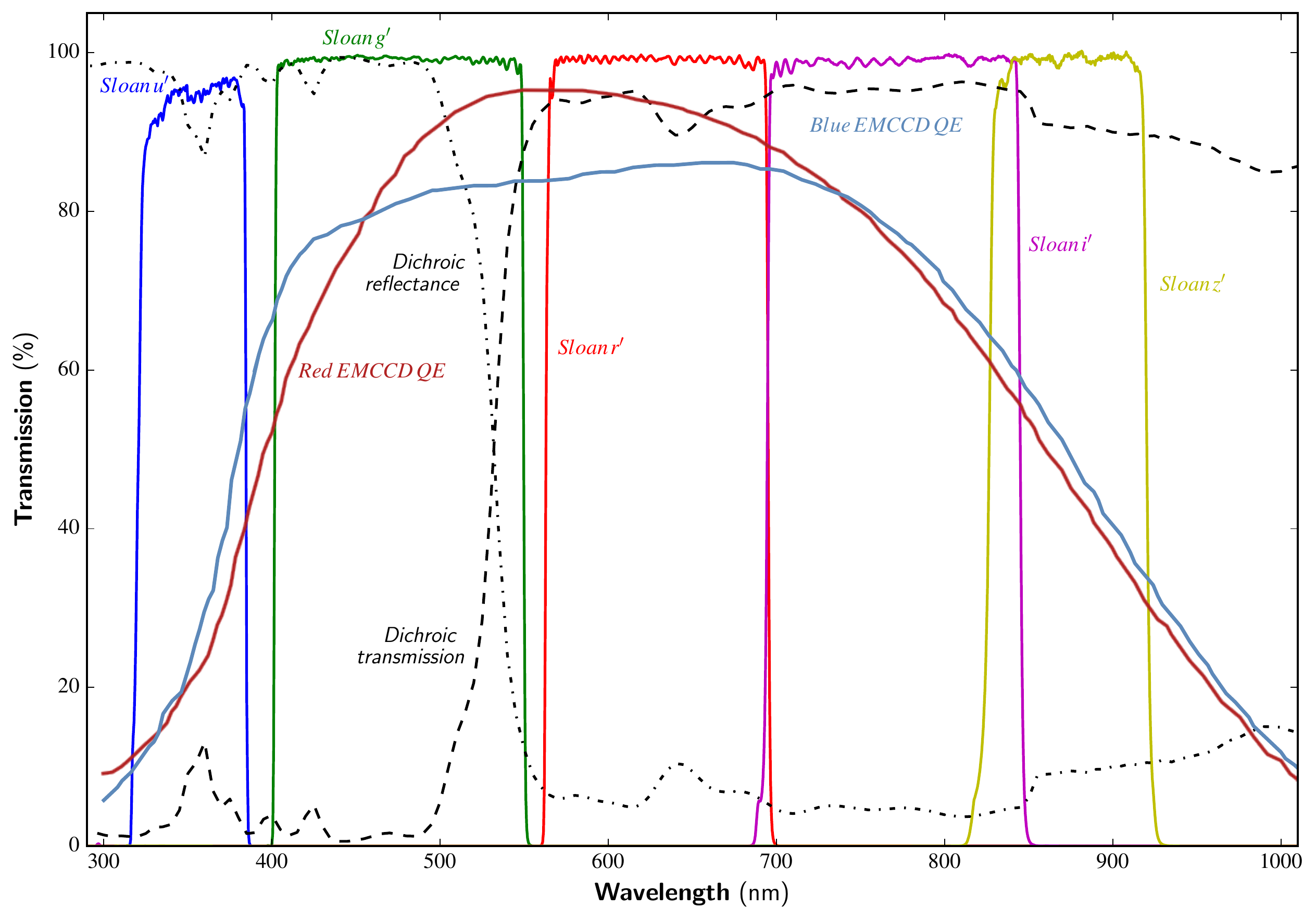}
\caption{CHIMERA filter and dichroic transmission curves, as well as the QE response of each detector.  The plane sensor window throughput is also integrated in to the QE curve calculation.  Filters curves reflect the Sloan u$^{\prime}$ or g$^{\prime}$ (blue camera), and Sloan r$^{\prime}$, i$^{\prime}$ or z\_s$^{\prime}$ (red camera).  The dichroic is centered at 567 nm, with a transmission-reflectance cross-over of 30 nm for $>$90\% response.  The blue camera sensor was coated with the e2v UV-vis coating, and the red camera sensor with the e2v vis-NIR coating.}\label{fig3}
\end{figure*}

As a result of the large dimensions in length and width, a thickness of 14 mm was specified in order to maintain glass stiffness to withstand coating and mounting stresses.  This was also important for transmitted wavefront accuracy to achieve $\lambda$/2 peak-to-valley across any 100 mm diameter circle, within the clear aperture.  Surface roughness is $<$2 nm rms, where typical values were reported to be 0.5 nm rms by the vendor.  Image quality was simulated in Zemax by adding both spherical, and astigmatism, to the dichroic surfaces.  We predicted that even with three waves of spherical plus astigmatism, the system would host most light in a two-by-two pixel region (two pixels $\sim$0.58 arcsec). We over-specified to one wave to meet this performance.

\subsection{Sloan broadband filters}

We designed a custom linear filter exchanger to house CHIMERA's filters, where each optical arm provided 4 available slots, see Figure~\ref{fig2}.  Sloan Digital Sky Survey (SDSS) filters \citep{fukugita96} were selected as the primary photometric filter set for CHIMERA, where the blue channel houses the Sloan u$^{\prime}$ and g$^{\prime}$ broadband filters, and the red channel, the Sloan r$^{\prime}$, i$^{\prime}$ and z\_s$^{\prime}$ filters.  This range matches the QE response of the CHIMERA detectors well, where the SDSS atmospheric cut-off at 300 nm lies approximately where the response of CHIMERA's silicon-based detectors approach zero.  This cut-off also applies to the z\_s$^{\prime}$ filter at $\sim$920 nm (Figure~\ref{fig3}).

CHIMERA's SDSS filter system was obtained from Astrodon\footnotemark[5], \footnotetext[5]{http://www.astrodon.com/}where we selected their SDSS `Gen2' filter variant.  The Gen2 offers additional spectral separation between the g$^{\prime}$ and r$^{\prime}$ filters, to better avoid atmospheric air glow where the (OI) 5577 \r{A} line occurs.  In addition, the z\_s$^{\prime}$ filter controls the high-wavelength cut-off, rather than the detector.  Out of band blocking is of the order $\leq$0.03\% in Gen2\footnotemark[5].  Each filter was specified at 110.0 $\times$ 110.0 $\pm$ 0.1 mm of striae-free fused silica, with a thickness of 3.0 $\pm$ 0.025 mm.  In order to allow sufficient mounting area in the filter exchangers, we specified a minimum clear aperture of 108.0 $\pm$ 0.1 mm, with blackened edges to minimize reflections.  Peak transmission was measured to be $>$95\% for all filters, and $>$90\% for u$^{\prime}$, with 1/4-wave propagated wavefront prior to coating in addition to \mbox{$<$0.5 arcmin} of substrate parallelism with surface roughnesses of \mbox{$1 - 1.5$ nm}.

The filter exchanger was designed such that the filter was not orthogonal to the optical beam, but tilted by 10$^{\circ}$ to prevent strong ghosting.  Narrowband filters are currently being considered to aid additional focused science objectives, e.g. H$\alpha$ or Na, with BW$\sim$3 nm.

\subsection{Re-imaging lenses}\label{lenses}

The Canon EF 200 \textit{f/2.0} ``double Gauss'' style lens was used as CHIMERA's re-imaging camera.  Like all double Gauss lens systems, the Canon EF 200 \textit{f/2.0} employs a virtual pupil on its stop located inside the lens itself.  In CHIMERA's design, the Canon EF 200 is driven by a real pupil located $\sim$100 mm in front of its entrance aperture, close to the dichroic.  Based on marketing material provided by \cite{canon}, the predicted image diameters are smaller than CHIMERA's single 13 $\mu$m pix; however, this material assumes that the lens will be driven as designed.  We simulated the performance of the Canon lens with the external (real) pupil and found that the Canon lens delivers most light into 5 $\mu$m over the entire FOV at visual wavelengths.

The Canon lens was simulated using Japanese patent `JP 2011, 253050, A'. Model glasses that match the refractive index, Abbe number and partial dispersions, are used in place of glass names.  The simulated image quality of the Canon lens shows image diameters that are nearly 15 $\mu$m, which far exceed the image quality of the Vixen and 200-inch optical systems.  The Canon lens thusly adds only a fraction of degraded image quality when compared to prior optics in the system.  The re-imaging camera has 15 surfaces with a total throughput of $\sim$85\% when averaged over $400 - 800$ nm.  The EF 200 expects an entrance pupil on its internal stop, whereas the 200-inch and Vixen collimator's exit pupil is presented $\sim$25 mm further downstream in the Vixen.  The vignetting is $\sim$10\% on-axis, to $\sim$40\% off-axis.  Since the pupils on the camera will move as a function of field position, where a pupil is located at the most extreme position, the fields are geometrically vignetted by the camera entrance aperture.  Therefore, the total geometric throughput can vary by between 90\% (on-axis), to 60\% at the edge of field.  The \textit{f/2.0} Canon lens provides a 1.9x demagnification, such that final plate scale is \mbox{$\sim$3.4 pix arcsec$^{-1}$}.

\subsection{EF232 focus/aperture control \& window}

Although the prime cage of the 200-inch offers a wide focus range for an instrument (75 mm via telescope control and an additional 125 mm mechanically), CHIMERA requires additional focus control on each channel to account for the distinct wavelength-dependent variation in focus in each band.  Therefore, we also included a remotely controlled focus and aperture controller, the EF-232, on each arm, that is placed in between the Canon EF 200 lens and the EMCCD sensor.  The EF-232 was obtained from Birger Engineering\footnotemark[7] \footnotetext[7]{http://www.birger.com/}and offers independent focus control for each channel.  In this way, the Canon EF 200 lens behaves as if it were connected to a Canon DSLR, and thus has full functionality.

\begin{table}
\caption{Theoretical throughput of the full optical system, including the 200-inch mirror and Wynne corrector, based on Zemax modeling.  The instrumental throughput alone is discussed in Section~\ref{performance}.  These models used the optical prescriptions taken from \citet{kells1998}, or provided by Edmund Optics, Vixen Optics, Custom Scientific, Inc., Astrodon, Canon Cameras.}
\begin{center}
\begin{tabular}{cccc}
\hline
Element & Geometric & Fresnel & Total \\
\hline 
200-inch+corrector & .91 & .85 & .77 \\
\hline
Vixen collimator & .85 & .88 & .75 \\
Dichroic & 1 & .90 & .90 \\
SDSS filter & 1 & .99 & .99 \\
Canon camera & .90 - .60 & .85 & .77 - .51 \\
Total CHIMERA & .76 - .51& .67 & .51 - .34 \\
\hline
Total (telescope+inst) & .70 - .46 & .55 &.38 - .26\\
\hline
\end{tabular}
\end{center}
\label{table1}
\end{table}

Finally, the last optical element of the CHIMERA collimator-camera system is the CCD window that acts to maintain the hermetically-sealed vacuum chamber containing the EMCCD sensor.  This is a plane, UV-grade fused silica window, approaching 99\% throughput.  See www.andor.com for more details.

\begin{figure}
\hspace{-3em}
   \includegraphics[width=10cm]{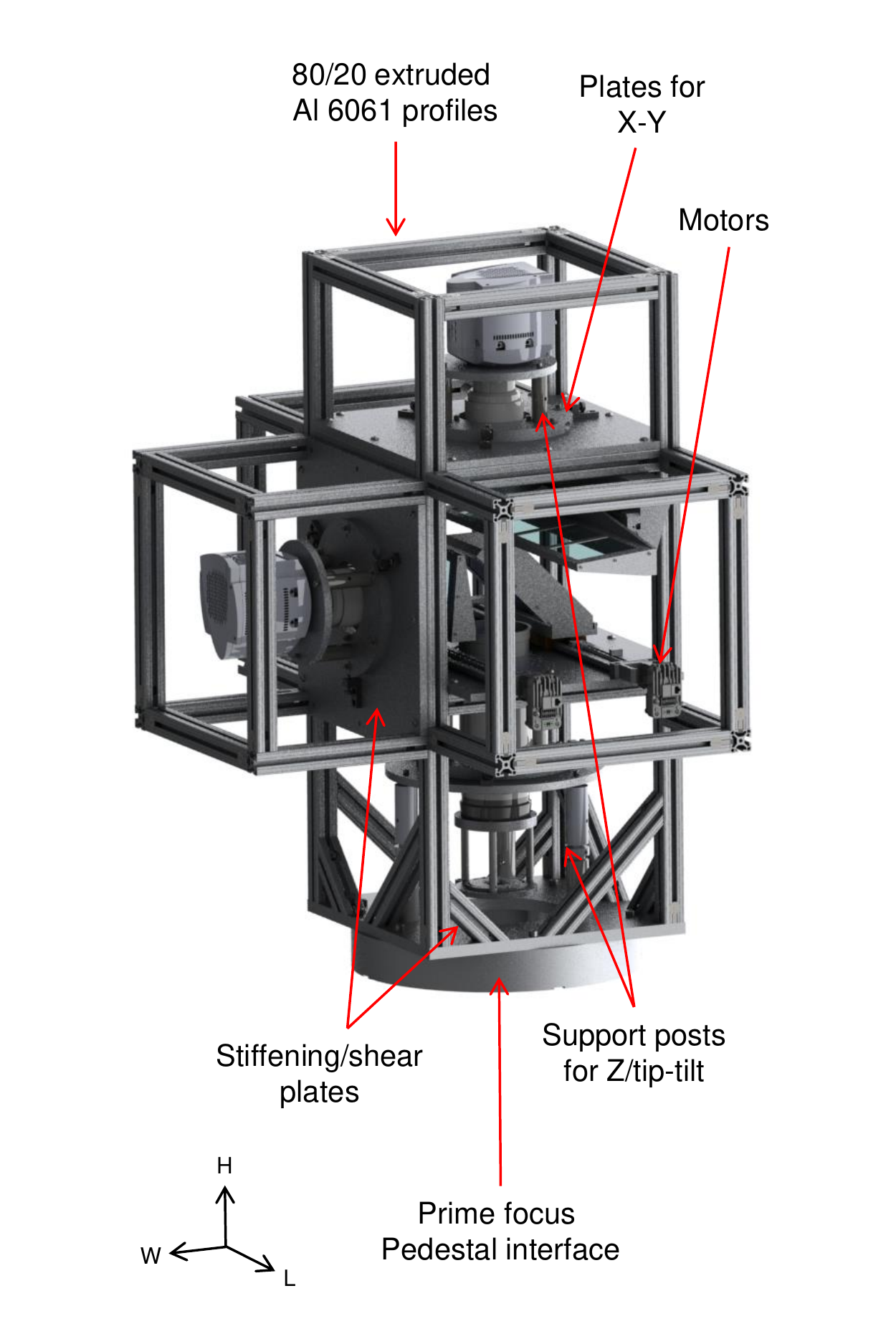}
\caption{3D CAD Solidworks render of the CHIMERA instrument.  We have removed the external paneling for an internal view of the optical design.  The main structure consists of 80/20 extruded Al 6061 profiles and 12.7 mm Al 6061 plates.  The collimator, dichroic beam splitter and re-imaging cameras, all have mounting designs that allow X, Y, Z and tip-tilt alignment.  The axes indicator on the bottom left indicates the H $\times$ L $\times$ W dimensions of 1.085 m, 0.968 m and 0.713 m, respectively.}\label{fig4}
\end{figure}

\section[]{Mechanical Design}
 
\subsection{Materials, structure \& custom opto-mechanics}

The support structure for the instrument was constructed out of precipitation-hardened 80/20 extruded Aluminium alloy profiles (Al 6061).  The main structure is a box-frame with diagonal supports at the base for strength and stiffness, as shown in Figure~\ref{fig4}, (\textit{left}), and facilitated easy access to optics and other components.  In addition to these supports, 12.7 mm Al 6061 plates were used to clamp sections of the instrument together; these also acted as stiffening plates.  We used \mbox{3.175 mm} anodized Al 6061 covers on the box-frame which provided shear support, and also aided light-tightness.  These components ensured structural stiffness during lateral loading at low telescope elevations.  We imposed a strict stiffness constraint such that the central beam would maintain its position to $<$1 pix ($<$13 $\mu$m) for all gravity vectors, including Palomar's lowest elevation angle cut-off of 18$^{\circ}$.

Newport and Thorlabs optical posts provided additional stiffness, where their primary function was to provide X, Y, Z and tip/tilt fine adjustment, for optical alignment -- see Figures~\ref{fig2} and~\ref{fig4}.  Custom mounts were designed and fabricated for the field lens and baffle, the Vixen collimator, the dichroic, the filters, and the re-imaging cameras (incl. Canon EF 200, EF-232 and EMCCD camera as one single part, hereafter re-imager).  In designing a single re-imager mount, this greatly helped optical alignment and ensured that once the prime focus central beam passed through the apex of the Canon EF 200, it would also be aligned with the geometric center of the EF-232 and hit the detector at the central X-Y pixel position, \textit{512, 512}.  As a result of the 100 mm filter optics, a custom 4-position linear exchanger and lead screw was also designed to facilitate movement over 4 positions, where custom mounting surfaces, limit/home switches and motors were integrated in to the design.

All materials were ordered as black, matted, or anodized where necessary.  All internal walls of the instrument were covered with black felt to minimize internal reflections.  We refer the reader to \citet{marshall14} for analysis on the characteristics of various materials that can reduce  spectral reflectance, where they show that black felt is an effective material at doing so.  The total mass of the instrument is $\sim$115 kg, including wiring and power supplies, 0.968 m in length and 0.713 m in width, see Figure~\ref{fig4}.

\subsection{Alignment}

We aligned CHIMERA by constructing an $\sim f/3.6$ optical system, which consisted of a collimated LED (470 -- 850 nm, 10 dB BW) and a lens ($D=25$ mm, $FL=90$ mm).  An LED driver with a drive current range of $200 -- 1200$ mA was used to control the intensity and neutral density filters were installed where necessary to avoid saturation.  We placed a 50 $\mu$m pinhole at the 200-inch focus point (113.26 mm above the prime focus pedestal interface).  All opto-mechanical elements were initially aligned to their Zemax model positions by using a FaroArm\footnotemark[8], \footnotetext[8]{See http://www.faro.com/home for more information}a portable coordinate mapping arm capable of high precisions in 3D metrology.  The prime focus pedestal interface was used as a zeroth reference point.

Once the system was collimated, we assessed the FWHM of the spot generated on the EMCCD after passing through the re-imaging optics.  The sensor's 13 $\mu$m pix pitch coupled with 50 $\mu$m pinhole yielded a FWHM of \mbox{$\sim$3.85 pix} when in focus, which was aligned to the sensor's central pixel.  Optical performance was compared to Zemax modeling thereafter.  Finally, the dichroic element was positioned in the central beam which we define as its zeroth position.  The motor's high resolution (4000 steps mm$^{-1}$) was encoded to ensure that the dichroic would always return to this position after being moved by the linear exchanger.

\section[]{Detectors: EMCCDs}\label{EMCCDs}

CHIMERA uses two Andor iXon Ultra 888 EMCCD cameras. These detectors house the CCD201-20 sensor from e2v, see Table~\ref{table2}.  This is a frame transfer (FT) EMCCD, with an active image section of 1024 $\times$ 1024 pix and a 13 $\mu$m pixel pitch.  The store, or FT section, has 1056 $\times$ 1037 pix (including dark reference rows) that allows high-duty cycle image acquisition, thus making the CCD201-20 the largest-format EMCCD device currently available from e2v, and ideal for CHIMERA's applications.

\begin{table}
\caption{Specifications of the CCD201-20 EMCCD, from e2v, as deployed in the Andor iXon Ultra 888 camera. $^{\star}$FT = Frame Transfer; $^{\ast}$BI = Back-Illuminated; $^{\ddag}$IMO = Inverted Mode Operation.  We also show some examples of the native read noise of some horizontal read out rates.}
\begin{center}
\begin{tabular}{cc}
\hline
Parameter & Specification \\
\hline 
Sensor type & EMCCD \\
Variant & FT$^{\star}$, BI$^{\ast}$, 2-phase \\
Active pixels (image) & 1024 $\times$ 1024 \\
Pixel pitch & 13 $\mu$m \\
Digitization & 16-bit \\
Amplifiers & Conv. or EM \\
Calibrated gain range & 1 - 1000 \\
Multiplication Stages & 604 \\
Read out rate (Conv. horiz.) & 1, 0.1 MHz \\
Read out rate (EM horiz.) & 30, 20, 10, 1 MHz \\
Frame rates & 26 fps (full) -- 1000 fps (sub) \\
Dark current (IMO$^{\ddag}$, -85$^{\circ}$ C) & 5 $\times$ 10$^{-4}$ e$^{-}$ pix$^{-1}$sec$^{-1}$\\
\hline
Read noise (Conv., 1 MHz) & $\sim$6 e$^{-}$ rms\\
Read noise (EM, 1 MHz) & $\sim$20 e$^{-}$ rms\\
Read noise (EM, 10 MHz) & $\sim$53 e$^{-}$ rms\\
Read noise (EM, 20 MHz) & $\sim$120 e$^{-}$ rms\\
Eff. read noise (w/EM gain) & $<$1 e$^{-}$ rms \\
\hline
\end{tabular}
\end{center}
\label{table2}
\end{table}

The CCD201-20 is a 2-phase, thinned, back-illuminated device and has dual output amplifiers: the conventional amplifier for high dynamic range and the EM amplifier for high-sensitivity.  Thinning and back-illumination helps to improve QE, which peaks  at $>$90\% at 550 nm.  The power of the EMCCD is in the EM process; however, this is inherently stochastic, where there is $\sim1-1.5\%$ probability ($\alpha$) of an extra electron getting generated per given amplification stage.  Since a device can contain hundreds of multiplication stages, the probability of amplification becomes significant\footnotemark[9]. \footnotetext[9]{$EM=(1+\alpha)^N$, where $N$ is the number of stages.  We note that there is an added variance in the EM output which effectively reduces the sensor's QE by up to $\sim$50\%. This is referred to as the ``Excess Noise Factor'' (ENF), and asymptotically approaches $\sqrt{2}$ for gains $\geq$10. Post-read out techniques can reclaim this reduction in QE, see \citet{daigle08}.} The high gain amplifier is driven by much higher voltages (\mbox{$>$40 V}) than the conventional register ($\sim$11 V), and can deliver an effective read noise of $<$1 e$^{-}$ rms with EM gain, whereas the conventional amplifier provides a noise of \mbox{$\sim$6 e$^{-}$ rms}.  Crucially, as a result of much higher bias voltages controlling the gain via the high voltage clock, relatively low amounts of raw signal can quickly lead to saturation of pixels in the high gain register.  Therefore, there is always a trade-off between the desired reduction in read noise and a reduction in the effective pixel charge capacity.  The Andor iXon Ultra 888 offers a calibrated gain range of $1 - 1000$, where the CCD201-20 can demonstrate a maximum well depth of 80,000 e$^{-}$ in the image section, and 730,000 e$^{-}$ in the gain register.  This is necessary to avoid saturation during the amplification process.  It is important not to continuously saturate the device under high amplification as this can cause irreparable damage to the EM register.

\begin{figure*}
   \includegraphics[width=17cm]{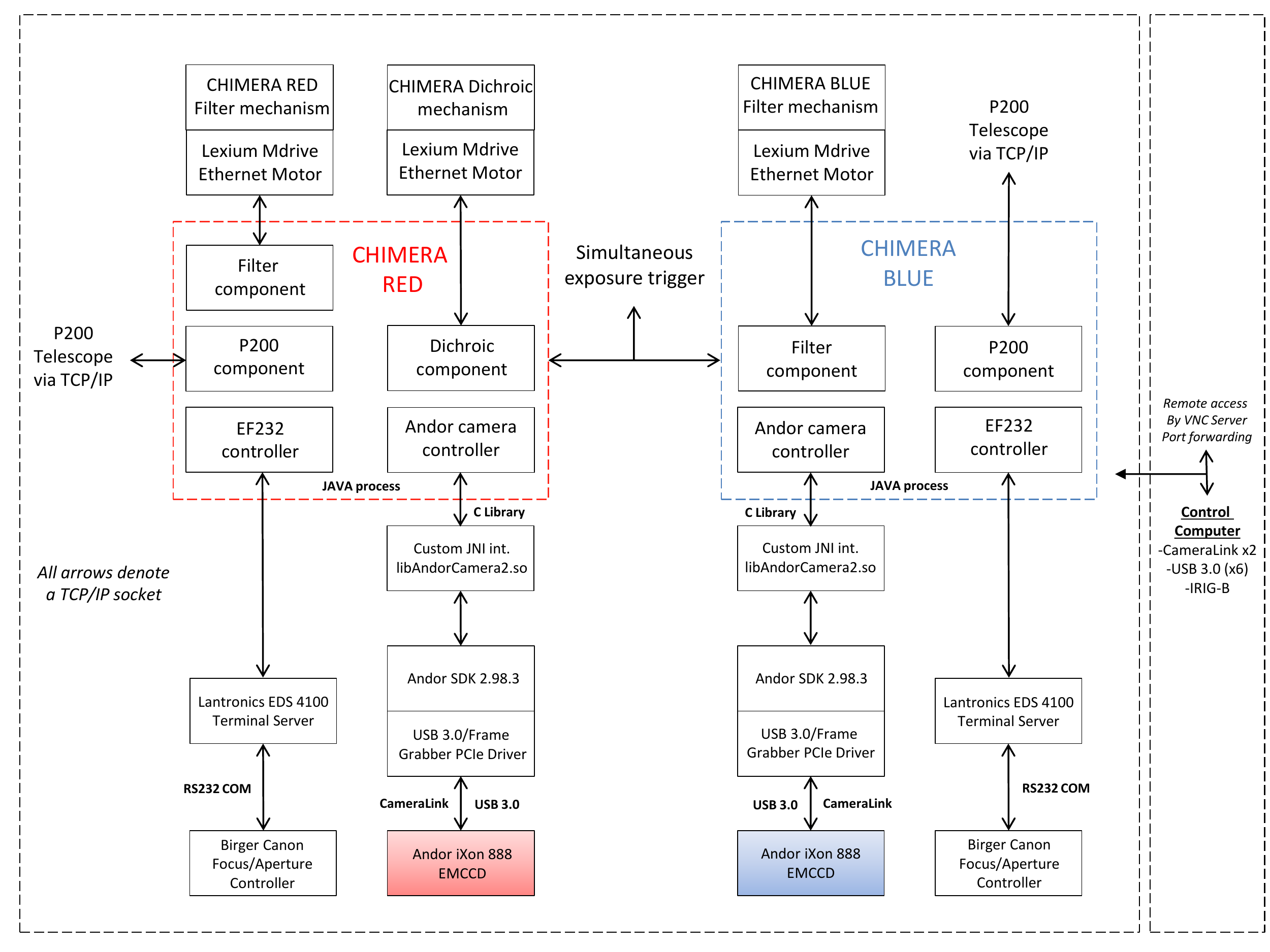}
\caption{Block diagram illustrating the CHIMERA data system and software communication logic.  The arrows represent a TCP/IP socket connection, where the large rectangular box on the left represents the full system which is controlled by the CHIMERA control computer, also located in the prime focus cage, shown in the small rectangular box to the right. }\label{fig5}
\end{figure*}

Charge is clocked out through the parallel (image and store section) and horizontal chains, where the iXon Ultra 888 offers variable parallel read out speeds of 0.6 -- 4.33 $\mu$s, and horizontal read out rates of 0.1, 1, 10, 20 and 30 MHz, allowing 1000s of fps.  The blue camera is coated with the e2v \textit{BUV} anti-reflective coating, whereas the red camera is coated with e2v's \textit{vis-NIR} anti-reflective coating, where both of the devices are science-grade sensors (e2v's highest cosmetic and performance yield).  

The devices are run in inverted mode operation (IMO), thus greatly suppressing generation of minority carriers - namely surface dark current.   Further dark current suppression is achieved by cooling the device to a stable -85$^{\circ}$ C by the Peltier effect via three-stage thermoelectric cooling (TEC).  However, we note that at higher frame rates (\mbox{$>$50 fps}), the TEC can only maintain stable temperatures at $\sim$ -65$^{\circ}$ C, yielding a $\sim$3 fold increase in dark current from \mbox{-85$^{\circ}$ C}.  This dark current generation is still sufficiently low for most ground-based applications \citep{dhillon07}.

\section[]{CHIMERA Software}\label{software}

\subsection{Control Software}
\subsubsection{Design \& integration with Andor} 

CHIMERA's control software uses a combination of Java (JDK 1.7) and C (gcc, nvcc) and relies on the `Netbeans' integrated development environment (IDE).  The software system makes extensive use of the JSky archive\footnotemark[10] \footnotetext[10]{http://archive.eso.org/cms/tools-documentation/jsky.html.}for image display and image quality analysis.  Ephemeris calculations are obtained from the JSkyCalc programs\footnotemark[11] \footnotetext[11]{Provided by John Thorstensen of Dartmouth College, http://www.dartmouth.edu/~physics/labs/skycalc/flyer.htm.l}where the software runs Red Hat Enterprise Linux (version 6.5).

The cameras are controlled by using the software development kit (SDK, v2.98.3) provided by Andor.  Since the software is compatible with any Andor EMCCD camera variant, the SDK can use either USB communication or a CameraLink frame grabber PCIe card for image acquisition.  Since the graphical user interface (GUI) is written in Java, a custom Java Native Interface (JNI) is required to access the Andor SDK.  Two frame grabber cards, each communicating with a single camera in the case of the iXon 888, are installed in a single computer.  Each instance of the software acts as a TCP/IP socket server and allows expose commands to be synchronized between the two cameras at $\sim$ms timescales.

\subsubsection{Data acquisition \& optics software control} 

The exposure process operates in a `run till abort' mode where the camera stores the latest image in an internal buffer ($\sim$90 full frame images), which is then retrieved by the software.  The camera internal buffer is treated as a FIFO (`first in, first out') buffer by checking the number of available images and retrieving the oldest image first. 

The dichroic and filter linear exchangers are driven by Lexium MDrive Ethernet NEMA 17 motors\footnotemark[13]\footnotetext[13]{Schneider Electric Motion, USA.}. Motion commands and requests for motor status are sent to the motors using MCode ASCII strings over the TCP/IP port for each mechanism and a custom control within the software allows each mechanism to be controlled from within camera control system.

The EF232 controller (aperture control of the Canon lens) is connected to a Lantronics EDS 4100 terminal server.  A command string (ASCII) is sent to the EF232 to query the mechanism state and to command focus and aperture position.  All camera and telescope parameters, the current state of the filter and dichroic mechanisms, as well as the focus and aperture state are automatically written into the FITS header.

\subsubsection{The control GUI \& 200-inch comms} 

The software runs in either `Targeting Mode', where all camera parameters may be modified while the system is acquiring images, or in `Science Mode', where the parameters are fixed at the beginning of a set of exposures.  Complete control of binning and region of interest allow the observer to operate in full frame, a set of pre-configured binning modes, and/or sub-framed modes.  It also supports custom binning.

The feature set support by the CHIMERA software is extremely extensive and therefore we only cover the main features here, but refer the reader to the `CHIMERA Users Manual' that can be found on the CHIMERA website for more information\footnotemark[14]\footnotetext[14]{http://www.tauceti.caltech.edu/chimera/}.  The software is fully integrated into the 200-inch telescope control system.  The telescope telemetry is polled at $\sim$10 Hz and as before, all relevant telescope position information is written into the FITS headers.  Telescope control is integrated into the image display system allowing observers to re-point the telescope by selecting positions on the image display, or by using simulated paddles.  A local copy of the UCAC3 catalog \citep{zacharias10} is integrated into the software and may be queried and displayed both in tabular form, and as an image overlay.

\subsubsection{Dithering \& guiding} 
 
A sophisticated dither and mapping control is included that allows observing patterns to be executed easily, e.g. 10 $\times$ 10 grid of the 5 $\times$ 5 FOV on sky, fully automated, including exposure triggering at each settle point and a halt at each slew.  The dither and mapping control contains an integrated visualization system that overlays the observing pattern on an SDSS image, and updates the state (by color) of the pattern during execution of the observation.  The CHIMERA software also supports high precision guiding using multiple guide objects based upon the code developed for planetary transit photometry with the WIRC instrument \citep{zhao12}.  The guide system uses multiple objects to separate motion due to seeing fluctuation from the mechanical drift of the telescope.  The system achieves guide stability of \mbox{$\pm$0.25 arcsec} ($<$1 pix) using this system.  We show a block diagram of the CHIMERA software in Figure~\ref{fig5}.

\begin{figure*}
   \includegraphics[width=17cm]{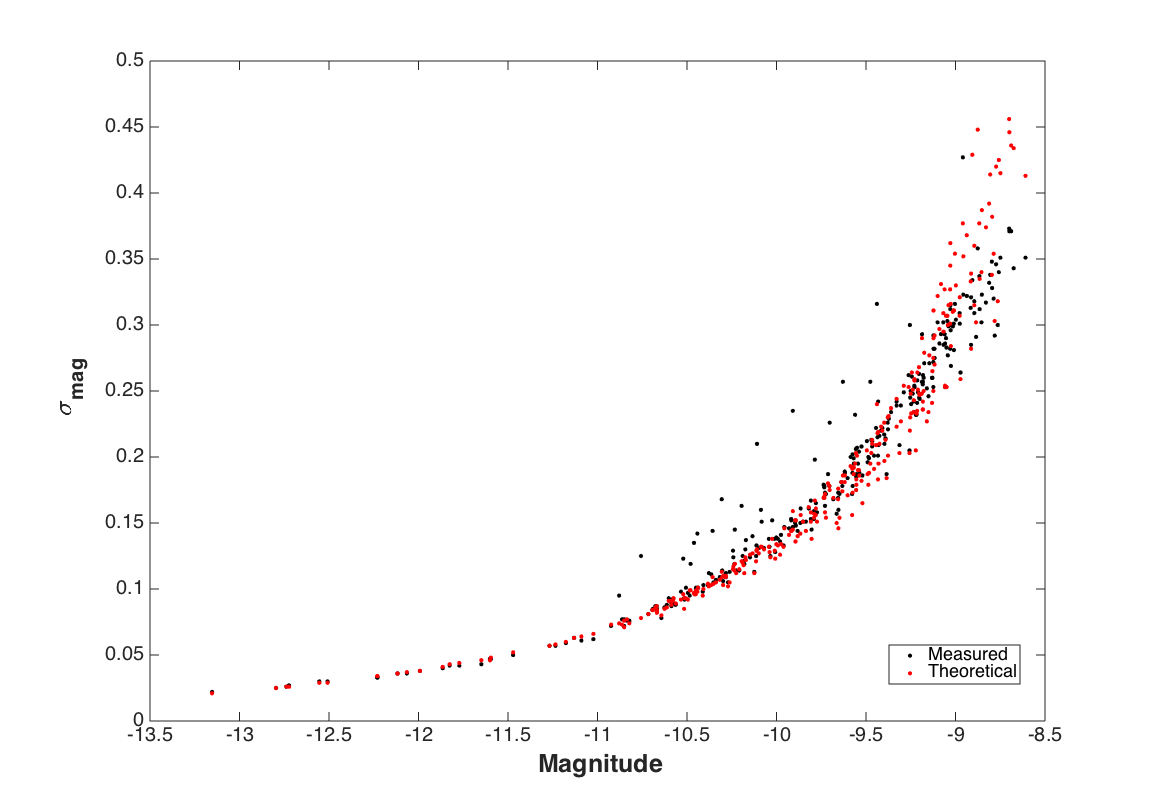}
\caption{$\sigma_{mag}$ vs. magnitude plot of the theoretical vs. measured noise performance of the CHIMERA instrument.  The instrumental magnitude of each star in the plot is $-2.5 \cdot log_{10} (flux)$.  The measured data (black data points) were obtained using the EM amplifier with an amplification gain of 100.  We used a KBO field at $\alpha=18:40:49.92$ and $\delta=-06:46:59.8$ since it provided hundreds of stars for analysis.  The following parameters were included in the calculation: photon noise, sky noise, read noise (with EM gain), dark current, clock induced charge (CIC), and the ENF.  See Equation~\ref{eq_phot} and Equation~\ref{eq_mag}, which show the methods used for this calculation.}\label{fig6}
\end{figure*}

\subsection{The CHIMERA Pipeline}

A data processing pipeline for the purpose of the reduction and processing of images from CHIMERA has been developed.  The three main functions of the pipeline are as follows: 1) to calibrate raw data, 2) to generate `first look' data products and 3) to provide detailed analyses tools for post-processing.  Image calibration involves bias subtraction, flat field correction, the flagging of bad pixels or columns and field distortion correction.  The pipeline offers routines to generate calibrated photometric light curves immediately after data acquisition is complete.  This serves as a useful tool for observers who wish to assess the behavior of an astronomical target during an observation window.  Since we consider this `first look' result as preliminary, the pipeline offers additional post-processing packages allowing a more detailed analysis of the raw data set.
  
The pipeline was developed in the Python environment using the \textit{Numpy}, \textit{Scipy} and \textit{Matplotlib} packages in order to provide cross-platform support.  Although the routines will not require the PyRAF environment to run, it can easily be run in conjunction with IRAF/PyRAF tasks.  This work made use of Astropy, a community-developed core Python package for Astronomy \citep{astropy13}, for source detection and aperture photometry.  Among other utility routines, the pipeline also includes routines to generate animation from data cubes, and phase folding of light curves.  An initial version of the pipeline has been released on the CHIMERA website (see footnote 14) as well as on the github\footnotemark[15]\footnotetext[15]{https://github.com/caltech-chimera/pychimera} version control system.   


\section[]{Instrument Performance}\label{performance}

\subsection{Noise Characterization}

Evaluating the noise in an EMCCD is similar to that of a conventional CCD; however, as noted in Section~\ref{EMCCDs}, if the high gain register under EM amplification is used, there is a gain factor applied which results in a reduction of the read noise, $\sigma_{RN}$, by an amount $R/G$, where $R$ is the number of electrons and $G$ is the amplification gain.  Additionally, the ENF produces an added variance in the output signal due to the stochastic nature of the EM process.  As a result, when calculating the S/N for an EMCCD, $\sigma_{RN}$ becomes negligable but the ENF and multiplication gain must be considered in addition to other paramters such as the clock induced charge (CIC) which can dominate under high gain conditions.  CIC is another source of spurious noise caused when a CCD is clocked into inversion, and since the Andor iXon 888 EMCCD camera is run in IMO to suppress surface dark current, CIC is greater than a device run in non-inverted mode (NIMO) and must be included.

In order to characterize CHIMERA's noise performance, which includes photon, sky and detector noise, we calculated the total theoretical noise of the instrument and compared this to the measured noise, which we show in Figure~\ref{fig6} as an $\sigma_{mag}$ vs. magnitude plot.  We note that the theoretical model and measured results were calculated/obtained using the EM amplifier, and included the following parameters: photon noise, sky noise, read noise (with EM gain), dark current, CIC, and the ENF.  We calculated the total theoretical S/N as follows:

\begin{equation}\label{eq_phot}
	S/N = \frac{N_{star} \cdot k}{\sqrt{\sigma_{source}^{2} + \sigma_{sky}^{2} + \sigma_{dark}^{2} + \sigma_{RN}^{2} + \sigma_{CIC}^{2}}}
\end{equation}
where:\\ \\
\noindent -- $N_{star}$ is the stellar flux in ADU.\\
\noindent -- $k$ is the conversion gain in $e^{-}$ ADU$^{-1}$.\\ 
\noindent -- $\sigma_{source} = \sqrt{ENF^{2} \cdot N_{star} \cdot k}$, which is the target shot noise.\\
\noindent -- $\sigma_{sky} = \sqrt{ENF^{2} \cdot N_{sky} \cdot k \cdot n_{pix}}$, which is the sky noise, where $N_{sky}$ is sky flux in photons and $n_{pix}$ is the number of pixels.\\
\noindent -- $\sigma_{dark} = \sqrt{ENF^{2} \cdot q_{dark} \cdot t_{exp} \cdot n_{pix}}$, which is dark current, where $q_{dark}$ is the dark current in e$^{-}$ pix$^{-1}$ sec$^{-1}$ and $t_{exp}$ is the integration time in seconds.\\
\noindent -- $\sigma_{RN} = \sqrt{(R/G)^{2} \cdot n_{pix}}$, which is the read noise.\\
\noindent -- $\sigma_{CIC} = \sqrt{ENF^{2} \cdot q_{CIC} \cdot n_{tr}}$, which is the CIC, where $q_{CIC}$ is the CIC per pixel and $n_{tr}$ is the number of transfers.\\

The measured instrumental magnitude of each star in Figure~\ref{fig6} is $-2.5 \cdot log_{10} (flux)$ (x-axis) and the standard deviation from the mean magnitude (y-axis) was found as follows:

\begin{equation}\label{eq_mag}
	\sigma_{mag} = 1.0857 \cdot N/S
\end{equation}

We obtained data from a moderately dense KBO field at $\alpha=18:40:49.92$ and $\delta=-06:46:59.8$ in order to show noise performance in the KBO high cadence regime.  These data were taken at 33 Hz in 2 $\times$ 2 binned mode on Aug 15, 2015, using the Sloan $i^{\prime}$ band, and were bias subtracted and flat fielded.  The data points in Figure~\ref{fig6} show 304 stars that were detected at 10$\sigma$ above background in an averaged image using the IRAF DAOPHOT \textit{daofind} routine  \citep{stetson87}, where  the theoretical performance of the system (calculated using Equation~\ref{eq_phot}) is shown as the red data points and the measured performance is shown as the black data points.  Dark current was measured to be \mbox{3 $\times$ 10$^{-4}$ e$^{-1}$ pix$^{-1}$ sec$^{-1}$} at -85$^{\circ}$ C and CIC was measured to be \mbox{8 $\times$ 10$^{-3}$ e$^{-1}$ pix$^{-1}$ transfer$^{-1}$} at a 20 MHz read out rate.  Since the read noise was found to be $\sim$130 e$^{-}$ rms at 20 MHz, we used an EM gain of 100 to achieve an effective read noise of $\sim$1.3 e$^{-}$ rms.  In this example, we used standard aperture photometry; however, for denser fields, PSF fitting should yield higher photometric precision since contamination from nearby sources can be avoided.

\begin{table*}
\caption{Empirically measured throughput of the CHIMERA optical system.  Note:  We use the AB magnitude system for CHIMERA's Sloan filter set.  The AB magnitude is when 0 magnitude stars have a flux of 3631 Jy in all bands; ZP = zero point.  We note that the throughput and ZP of the Sloan u$^{\prime}$ band will be obtained during an upcoming CHIMERA observation.  We do not include current estimates due to data calibration issues.
}
\begin{center}
\begin{tabular}{ccccccc}
\hline
Filter & $\lambda$-central & Bandwidth & Theoretical AB ZP & Measured AB ZP & Theoretical throughput & Measured throughput \\
 & (nm) & (nm) & (mag) & (mag) & (\%) & (\%)\\
\hline 
u$^{\prime}$ & 352.5 & 65 & 25.56 & 23.63 & 9.40& 1.60\\
g$^{\prime}$ & 475.5 & 149 & 27.57 & 27.36  & 35.30 &29.15  \\
r$^{\prime}$ & 628.5 & 133 & 27.22&27.11 &38.00 & 34.13  \\
i$^{\prime}$ & 769.5 & 149 & 26.93 & 26.46 &31.70 &20.47  \\
z\_s$^{\prime}$ & 873.0 & 94 & 25.70 & 24.55  & 18.40&6.36  \\
\hline
\end{tabular}

\end{center}
\label{table3}
\end{table*}

Finally, we refer the reader to \citet{levitan14} for an example of CHIMERA's noise performance in a slower time domain, where they used the conventional amplifier to obtain a light curve of an AM CVn system ($M_{g^{\prime}}$ = 20.1) with 5 second integration times in the Sloan g$^{\prime}$.  Additionally, \citet{harding15} have recently observed an eclipsing binary ($M_{g^{\prime}}$ = 13.57) and found an occultation event with a depth of 0.07945 $\pm$ 0.00092 \% in g$^{\prime}$ and 0.029876 $\pm$ 0.00058 \% in r$^{\prime}$.  They also used the conventional amplifier with integration times of $\sim$5 seconds.  Both of these examples exhibited similar CIC and dark current (data taken at -85$^{\circ}$ C), and can therefore be considered representative of CHIMERA's performance using the conventional amplifier.

\subsection{Throughput \& Zero points}

We determined the optical throughput and standard photometric zero points (ZP) of the instrument in each of the Sloan bands.  Since the throughput of the system is dependent upon the transmission of the telescope and Wynne corrector, the collimator, the dichroic, the filter, the Canon lens and the QE of the sensor, the theoretical throughput for each filter was calculated by multiplying the transmission of each of these components.

We used the standard AB magnitude system, $m_{AB}$, to determine ZPs for each band.  The AB system is defined as the logarithm of the spectral flux density for a ZP of \mbox{3631 Jy}:

\begin{equation}
   m_{AB} = -2.5 \cdot log_{10}(F_{\nu}/3631 Jy),
\end{equation}
where $F_{\nu}$ is the spectral flux density and 1 Jy is 10$^{-26}$ $W m^{-2} Hz^{-1}$.  Therefore:

\begin{equation}
m_{AB} = -2.5 \cdot log_{10}(F_{\nu}) - 56.1
\end{equation}

The standard photometric star Hilt190 ($\alpha=01:58:24.07$ and $\delta=+61:53:43.5$) was used for all throughput calculations.  These data were bias subtracted and flat fielded and aperture photometry was carried out using the DAOPHOT task \textit{phot}.  We used the curve of growth method to determine a nominal aperture and found that an aperture of 8 pix yielded the highest S/N.  The flux (ADU) was converted to photons by first scaling by the conversion gain (e$^{-}$ADU$^{-1}$), followed by the QE of the sensor.  The Sloan u$^{\prime}$, g$^{\prime}$, r$^{\prime}$, i$^{\prime}$, z$^{\prime}$ magnitudes of Hilt190 were taken from \citet{smith02}.  The quotient of the measured photon counts and estimated photons counts from Hilt190 was used to estimate the instrument throughput. 

\begin{figure*}
   \includegraphics[width=17cm]{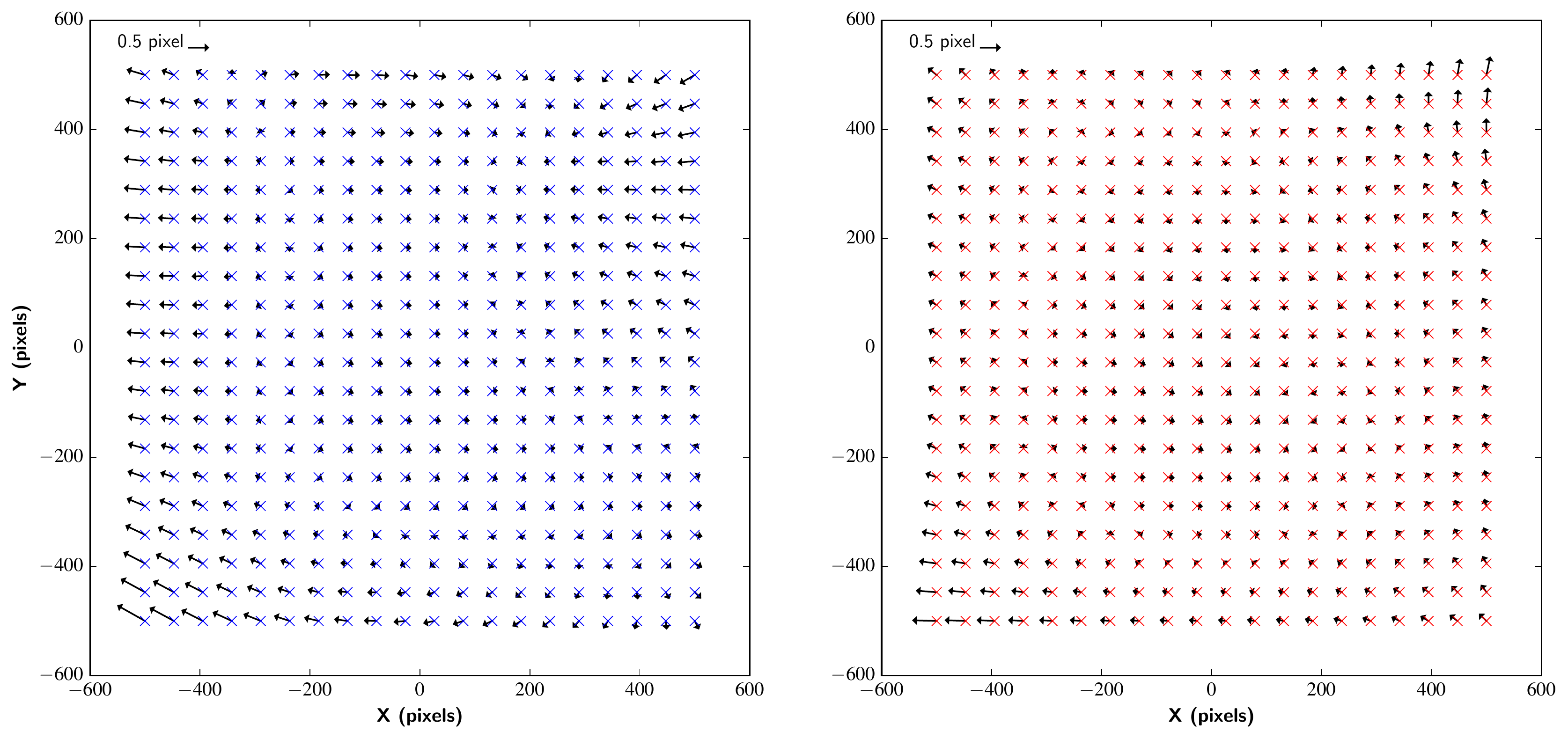}
\caption{Third order polynomial solution of the CHIMERA field distortion. Residuals of the fit are plotted as vectors for a 20 $\times$ 20 grid of points for both the blue and red optical channels.  The blue and red crosses signify actual grid position, and vector magnitude and direction show distortion at that point.  The coordinate system origin is at the center of the chip (512, 512).  The pixel scale $\sim$0.3 arcsec pix$^{-1}$ for CHIMERA's 13$\mu$m pitch, and the relative size of each pixel is shown in the top left of each plot.  We note that the scaling has been modified showing a residual vector zoom of 60$\times$ to give the reader a clearer understanding of distortion in the center.}\label{fig7}
\end{figure*}

The expected ZP was calculated by first establishing the number of photons sec$^{-1}$ that arrive at the telescope from a magnitude 0 star, and then multiplying by the throughput.  We assumed an atmospheric transmission of 1.0.  The instrumental magnitude of Hilt190 was calculated using the following equation:

\begin{equation}
m_{inst} = -2.5 \cdot log_{10}(f_{c}/t_{exp}),
\end{equation}
where $f_{c}$ is counts and $t_{exp}$ is exposure time.  Since this calculation reflects ground-based data, we therefore correct for atmospheric effects before calculating the instrumental ZP.  Atmospheric extinction effects were estimated using Palomar extinction curves given in \citet{hayes75}, and the extinction coefficient was multiplied by the airmass of Hilt190.  The above atmospheric instrumental magnitude is therefore given by:

\begin{equation}
m_{inst} = m_{inst0} + y \cdot X,
\end{equation}
where $m_{inst0}$ is the instrumental magnitude above the atmosphere, \textit{y} is the extinction coefficient and \textit{X} is the airmass. The measured ZP in the AB magnitude system is then found as follows:

\begin{equation}
m_{ZP} = m_{BD} - m_{inst0hilt},
\end{equation}
where $m_{BD}$ is the magnitude of Hilt190 in the AB magnitude system and $m_{inst0hilt}$ is instrumental magnitude of Hilt190 above the atmosphere.  The expected throughputs and ZPs, as well as measured throughputs and ZPs, are listed in Table~\ref{table3}.

We note that the Sloan u$^{\prime}$,  i$^{\prime}$ and  z\_s$^{\prime}$ measured throughput values are lower than what we predicted.  We used the theoretical throughput values from Table~\ref{table1} for our predictions and are confident that the performance for the telescope and Wynne corrector, collimator, dichroic and filters are all accurate, since we obtained empirically measured throughput data from each vendor.  However, the Canon lens throughput trace is propreitary data and was not provided by Canon.  Since the simulated Japanese patent (JP 2011, 253050, A) Zemax file only provided simulated data in the $\sim$400 -- 800 nm range (see Section~\ref{lenses}), thus excluding the u$^{\prime}$ and  z\_s$^{\prime}$ bands and partically excluding the i$^{\prime}$ band, we conducted a laboratory experiment that was designed to measure the true throughput of the lens at \mbox{$<$400 nm} and \mbox{$>$800 nm} wavelengths.  There was a large amount of uncertainty associated with this measurement, where we were unable to accurately measure the throughput in u$^{\prime}$ and z\_s$^{\prime}$, and found an approximate measurement in the 15 -- 20\% range in i$^{\prime}$.  Therefore we believe that our measured throughput, as shown in Table~\ref{table3}, is representative of CHIMERA's performance, since the predicted Canon throughput values are likely overestimated at these wavelengths.

\subsection{Field Distortion}

All optical systems suffer from aberrations and field distortions of some kind.  The mapping of these distortions and subsequent correction is essential for high precision photometry and astrometry.  This is especially important for the detection of NEAs, as outlined previously in this document.  We have performed distortion mapping of both the blue and red channels; however, we note that for all NEA observations, we remove the dichroic beam splitter and only use the red camera (over its full QE range) where optical distortions are lower. 

Images of the globular cluster M15 were taken using the Sloan i$^{\prime}$ filter and used to determine field distortion of the red camera.  All data were bias subtraction and flat field corrected prior to distortion correction; the \textit{daofind} task, was used for source extraction.  Higher proper motion stars were exluded from the sample in order to reduce the rms in the astrometric solution.  Those stars that were detected were matched thereafter with the UCAC4 catalog\footnotemark[16]\footnotetext[16]{http://dc.zah.uni-heidelberg.de/ucac4/q/s/form}.  A linear solution between catalog positions and measured pixel values was used to determine a plate solution.  A third order polynomial was carried out on these data to determine the field distortion by fitting the residuals from the plate solution.  A total of 320 stars were used to fit this polynomial.  The rms of the residuals was $\sim$70 mas.  A similar procedure was performed for the blue optical path, where we used the open cluster NGC 2158 (during observations of M15 for the red channel, the blue camera was being used for another calibration test). The rms of the fit was $\sim$100 mas.  We note that NGC 2158 is a much sparser field than M15, and this, in addition to chromatic effects from bluer stars and the presence of high proper motion field stars may all have contributed to a reduced astronomic accuracy solution for the blue camera.  We show this analysis in Figure~\ref{fig7}.

CHIMERA's distortion solution will be further improved by using dithered images of a globular cluster in upcoming observations.  Once a solution of $\leq$50 mas is achieved, the system's astrometric accuracy as calculated above will become limited by UCAC4 catalog errors, which are \mbox{$\sim$50 mas} for its brightest stars.  The upcoming GAIA catalog will improve such precision even further, providing positional accuracy of $\sim$10 mas.  Updated distortion correction information will be available on the CHIMERA website.

\subsection{Timing}\label{timing}

The Andor iXon Ultra 888 EMCCD camera contains an internal buffer that can hold a total of 97 full frame images, where the timing accuracy of image acquisition is dependent upon the FPGA clock, accurate to $\sim$1 ns.  The data rate is limited by the read out speed of the camera.  Since CHIMERA is operated by a standard version of Linux (RHEL 6.4), absolute time is ultimately dependent upon the system clock, which has been synchronized with the NTP (network time protocol) service at Palomar.  To guarantee absolute timing over different epochs, we use a Meinberg TCR170PEX GPS card that is synchronized with an IRIG-B broadcast signal.  Absolute timestamps are written into the FITS header with a accuracy and precision of 1 ms during default observing conditions.  Sub-ms timing is possible, but requires an external trigger using a TTL signal, which can be synced to the GPS timing card.

\begin{figure}
   \includegraphics[width=8.5cm]{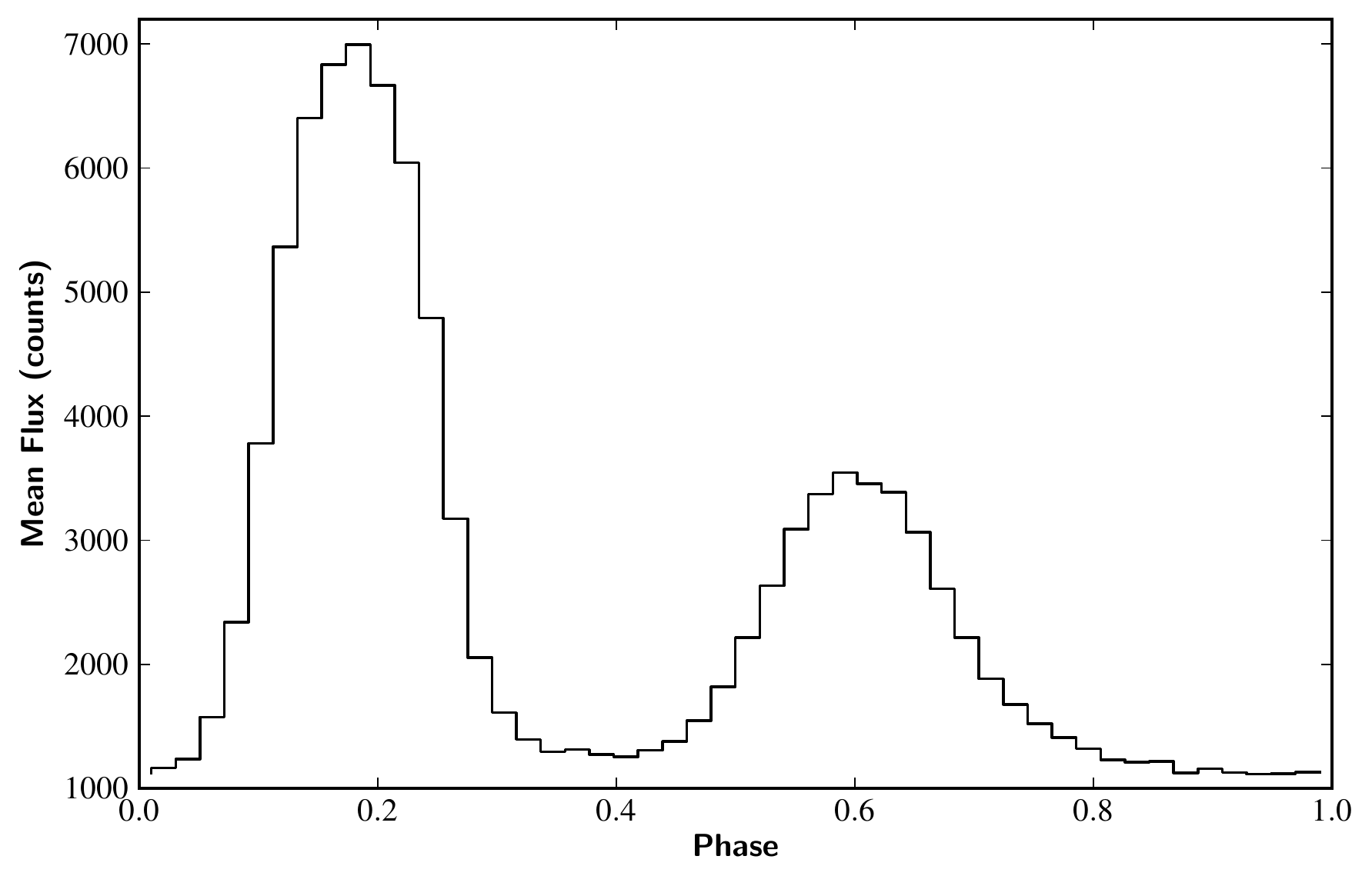}
\caption{Crab pulsar phase folded light curve. The main peak reflects pulsations at 29.595 Hz and is followed by an interpulse, see \citet{shearer03}.  Note that the pulses shown here are much broader than the pulsar pulse intrinsic width from \citet{shearer03}.  This is likely due to insufficient time resolution since our data was taken with an effective exposure time of 150 Hz.  To further increase the accuracy, a rate of $>$500 Hz would be more suitable.}\label{fig8}
\end{figure}

CHIMERA's timing accuracy was tested on-sky on Nov, 29, 2014 UT.  Windowed (64 $\times$ 64 pix), 2 $\times$ 2 binned images, were taken at 4.9 ms of the Crab Pulsar yielding an effective exposure time of 6.7 ms (including charge transfer).  Each data cube consisted of 20,000 frames.  The \textit{phot} task was used to extract the pulsar flux from each image and the resultant light curve analysis relied on the timing accuracy in order to phase fold the data to the expected Crab pulsation frequency.  We estimated this to be 33.7 $\pm$ 1 ms.  By comparing this directly to the Crab pulsar ephemeris of 33.792 ms from the Jodrell Bank Center for Astrophysics, November, 2014\footnotemark[17]\footnotetext[17]{http://www.jb.man.ac.uk/pulsar/crab.html}, we have demonstrated that the CHIMERA timing is accurate to the millisecond based on the system described above.  We show the phase folded light curve in Figure~\ref{fig8} and the pulsar is shown in the center of Figure~\ref{fig9}, (\textit{left}).

\section[]{First Light \& Future Upgrade}\label{section_future}

\subsection{First Light}\label{first_light}

\begin{figure*}
   \includegraphics[width=17cm]{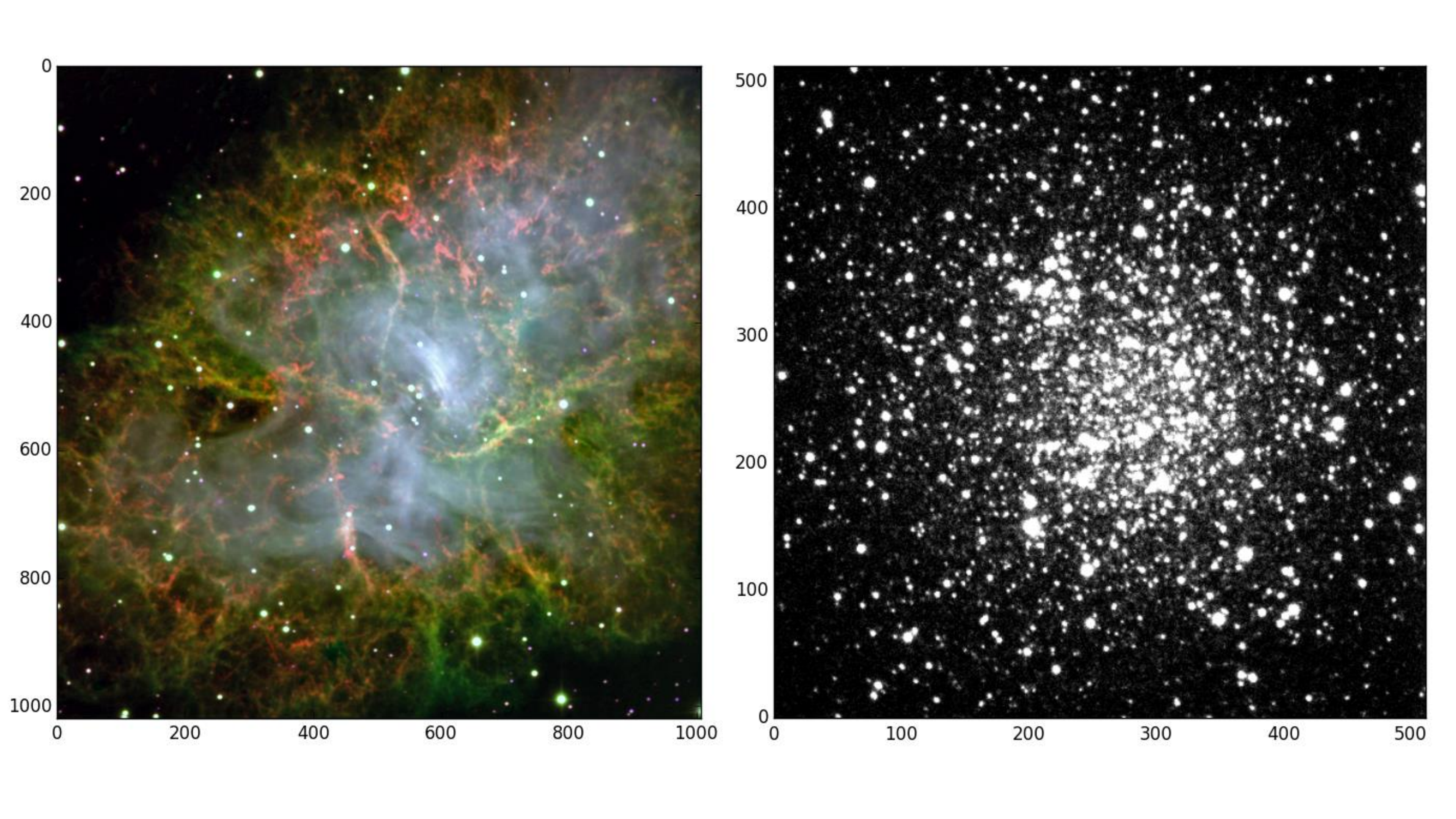}
\caption{CHIMERA first light images.  The axes of each image are in Pixels.  We show these images to demonstrate the versatility of CHIMERA in vastly different time domains where it is capable of obtaining deep pointings of extended objects or extremely high-cadence images of point sources and densely crowded fields.  (\textit{Left}) g$^{\prime}$, r$^{\prime}$, i$^{\prime}$, 300 second exposure of M1, the Crab Nebula.  Note the shock fronts that CHIMERA has detected that are emanating from the Crab pulsar at the center of the image. (\textit{Right}) A single 25 millisecond image with CHIMERA of the M22 field at 1$^{\circ}$ ecliptic latitude.  This frame contains $>$1000 stars with S/N$>$10, and was binned 2 $\times$ 2 to achieve the required cadence of 40 fps for detecting smaller-mass KBOs over the full 5 $\times$ 5 arcmin FOV.}\label{fig9}
\end{figure*}

CHIMERA saw first light in July, 2014.  We had a total of four nights, spread out between July 19, 24, 30 and 31 PST.  During this campaign, we carried out the necessary observations in order to fully characterize the instrument's performance, as described in Section ~\ref{performance}, as well as KBO and NEA-related science demonstrations.  We show two of CHIMERA's first light images in Figure~\ref{fig9}: M1, the Crab Nebula (\textit{left}) and M22, a globular cluster (\textit{right}).  The M1 image is a 300 second combined exposure in g$^{\prime}$, r$^{\prime}$ and i$^{\prime}$, capturing the full nebula over CHIMERA's 5 $\times$ 5 FOV.  In subsequent windowed observations, where data was used for timing accuracy calculations (Section~\ref{timing}), we detected individual pulses from the Crab pulsar that were successfully phased together, which pulses at approximately 30 Hz.  CHIMERA also detected shock fronts from the Crab Pulsar, as shown in the center of the image.  The M22 image is a single shot, taken at 25 ms, where CHIMERA detected $>$1000 stars with S/N$>$10.  This was taken for distortion testing and to test a potential KBO field at a 1$^{\circ}$ ecliptic latitude.  The disparity of operating modes and subsequent optical performance in these two examples is a clear example of CHIMERA's versatility as an astronomical instrument.

A prototype version of CHIMERA, with a narrower field of view, was tested in 2013 and has been used to observe a range of astronomical targets, including the most sensitive photometric detection of an AM Canum Venaticorum (AM CVn) in outburst, and detection of the faintest NEA (\mbox{H $=28$ mag}) yet discovered in the solar system \citep{levitan14,chengxing14,shao14,harding15}.  The wide field capability of the fully developed CHIMERA instrument can now be used to address the primary science goal of searching for KBO occultations.

\subsection[]{CHIMERA Future Upgrade}

The fully commissioned instrument presented in this paper was designed following a generation 1 ``prototype'', that was built in 2012, and tested and used in 2013.  CHIMERA's prototype, the Mark I, was designed as a proof-of-concept instrument.  The Mark I was designed with standard \mbox{50.8 mm} COTS optics, that provided a $\sim$2.5 $\times$ 2.5 arcmin FOV; however, optical distortion was present over much of its field.  

CHIMERA is on a path to installation as a facility instrument at the Hale telescope.  However, there is a planned substantial `Generation 3' upgrade, which will aim to image the fully-corrected prime focus FOV of \mbox{$\sim$25 arcmin $\times$ 25 arcmin} in two optical colors simultaneously.  In addition to this, a custom larger format sensor ($>$4K $\times$ 4K) will be developed, that will minimize optical components necessary to capture the full prime beam, and furthermore will provide $>$50 Hz frame rates over its full image section with rates of $>$1000 Hz on windowed sections of the sensor.  We note that the prime focus can offer a \mbox{$\sim$0.5$^{\circ}$ $\times$ 0.5$^{\circ}$} uncorrected FOV; however, large-format custom optics would be required to correct for optical distortion - this will also be considered.  This development will begin in parallel to larger EMCCD or CMOS sensor development -- a recent paper by \citet{gach14} describe a 4K $\times$ 4K EMCCD in the early stages of development but large-scale fabrication timelines are currently unknown.

\section[]{Conclusion}

We have presented a new high-speed, multi-color, wide-field photometer, CHIMERA, that will operate at the prime focus of the Hale 200-inch telescope.  It has been developed to specifically search for KBOs and together with the 200-inch telescope, offers a novel approach to detect these objects.  The design and commissioning of CHIMERA was a significant challenge, based on the optical aberrations at prime focus and the tight opto-mechanical constraints imposed by its collimator-camera design.  CHIMERA was fully commissioned in July, 2014, and its primary KBO campaign is now underway.  Additional key science carried out by CHIMERA will include searching for near Earth asteroids, in addition to monitoring short duration transient and periodic sources, including those discovered by the intermediate Palomar Transient Factory (iPTF), and the upcoming Zwicky Transient Facility (ZTF).  A future upgrade of CHIMERA is planned that will extend its current 5 arcmin $\times$ 5 arcmin FOV to 25 arcmin $\times$ 25 arcmin, thereby imaging the fully-corrected focal plane of the Hale 200-inch at $>$50 Hz full frame.  This upgrade is extensive and is at the early stages of conceptual design.

\section*{Acknowledgments}

The authors gratefully acknowledge the support of the Caltech Optical Observatories. We would especially like to thank Richard Dekaney and Christoph Baranec, for their extremely helpful advice throughout the CHIMERA project. We would like to highlight the excellent support of the Palomar Observatory staff, particularly John Henning, Steve Kunsman, Mike Doyle, Kevin Rykoski, Bruce Baker, Jamey Eriksen, Carolyn Heffner, Dan McKenna, Jean Mueller, Kajsa Peffer and Greg Van Idsinga. We also acknowledge Steve Macenka and James McGuire for their helpful contribution.  The work described here was carried out at the California Institute of Technology and the Jet Propulsion Laboratory (JPL), California Institute of Technology, under a JPL R\&TD grant and a contract with the National Aeronautics and Space Administration.\\\\




\bsp

\label{lastpage}

\end{document}